\input harvmac.tex
\noblackbox
\input epsf.tex
\overfullrule=0mm
\newcount\figno
\figno=0
\def\fig#1#2#3{
\par\begingroup\parindent=0pt\leftskip=1cm\rightskip=1cm\parindent=0pt 
\baselineskip=11pt
\global\advance\figno by 1
\midinsert
\epsfxsize=#3
\centerline{\epsfbox{#2}}
{\bf Fig. \the\figno:} #1\par
\endinsert\endgroup\par
}
\def\figlabel#1{\xdef#1{\the\figno}}
\def\encadremath#1{\vbox{\hrule\hbox{\vrule\kern8pt\vbox{\kern8pt
\hbox{$\displaystyle #1$}\kern8pt}
\kern8pt\vrule}\hrule}}
%%%%%%%%%%%%%%%%%%%%%%%%%%%%%%%%%%%%%%%%

\lref\Paul{P. Fendley, hep-th/9804108, to appear in
Advances in Theoretical and Mathematical Physics.}
\lref\FS{P. Fendley, H. Saleur, Phys. Rev. Lett. 81 (1998)
2518, cond-mat/9804173}
\lref\FSii{P. Fendley and H. Saleur, ``Hyperelliptic
curves for multi-channel quantum wires and the multi-channel Kondo
problem'', cond-mat/9809259}
\lref\SW{N. Seiberg, E. Witten, Nucl. Phys. B426 (1994), 19; Nucl. Phys.
B431 (1994) 484.}
\lref\FLSbig{P. Fendley, A.W.W. Ludwig and H. Saleur, Phys.
Rev. B52 (1995) 8934, cond-mat/9503172}
\lref\BLZ{V. V. Bazhanov,
S. Lukyanov, and A. B. Zamolodchikov, Comm. Math. Phys. 177 (1996)
381, hep-th/9412229; Comm. Math. Phys.
190 (1997) 247, hep-th/9604044; Comm. Math.
Phys. 200 (1999) 297, hep-th/9805008; hep-th/9812091.}
\lref\FLeSii{P. Fendley, F. Lesage, H. Saleur,
J. Stat. Phys. 85 (1996) 211, cond-mat/9510055.}
\lref\LeS{F. Lesage and H. Saleur, ``Perturbation of IR fixed points
and duality in quantum impurity problems'', cond-mat/9812045.}
\lref\Kondorev{A.M. Tsvelick and P.B. Wiegmann, Adv. Phys. 32 (1983)
453; N. Andrei, K. Furuya and J. Lowenstein, Rev.
Mod. Phys. 55 (1983) 331; N. Andrei and P. Zinn-Justin,
Nucl. Phys. B528 (1998) 648, cond-mat/9801158.}
\lref\FW{V. Fateev and P.B. Wiegmann, Phys. Lett. 81A (1981) 179}

\Title{}
{\vbox{
\centerline{Differential equations and duality }
 \vskip 4pt
\centerline{in massless integrable field theories at zero temperature}}}

\centerline{P. Fendley$^1$ and H. Saleur$^2$}
\bigskip\centerline{$^1$Department of Physics}
\centerline{University of Virginia}
\centerline{Charlottesville, VA 22901}
\bigskip\centerline{$^2$Department of Physics}
\centerline{University of Southern California}
\centerline{Los Angeles, CA 90089-0484}

\vskip .3in

Functional relations play a key role in the study of integrable
models.  We argue in this paper that for massless field theories at
zero temperature, these relations can in fact be interpreted as
monodromy relations. Combined with a recently discovered duality, this
gives a way to bypass the Bethe ansatz, and compute directly physical
quantities as solutions of a linear differential equation, or as
integrals over a hyperelliptic curve. We illustrate these ideas in
details in the case of the $c=1$ theory, and the associated boundary
sine-Gordon model.

\Date{3/99}

Integrable models are one of the most widely studied areas of
mathematical physics. The fact that there are an infinite number of
conserved charges commuting with the Hamiltonian means that
in principle physical quantities can be computed exactly. In practice,
this task is usually quite difficult.

Indeed, there are surprisingly few techniques at our disposal.
One of the most successful
ones is actually a related set of techniques going by the name of the
Bethe ansatz. In situations where the Bethe ansatz is applicable, one
makes some technical assumptions about the wave function, and then can
derive a set of equations for the zeroes of the wavefunction. The
number of such zeroes goes to infinity in the continuum limit, and one
can then use various density functions and the integral
equations they satisfy. The equations obtained are rarely solvable in
closed form but a great deal of quantitative information can be gained
by studying them in various limits or by numerical solution.

Since the Bethe ansatz is usually technically formidable and often
intractable, it is quite desirable to develop an alternative
formulation. One method being explored in recent years is the use of
functional relations or ``fusion'' relations for physical
quantities. The existence of such relations is almost obvious, and
based on very general properties of the Yang-Baxter equation and
quantum groups.  The assumptions made generally involve various
analyticity properties of physical quantities, and can often be
verified using the other approaches. Quite frequently, it can be shown
that the Bethe ansatz equations follow from the fusion relations, but 
the latter seem more powerful, more convenient, or deeper \BLZ. 

Another method that appears promising is the use of {\sl duality}.  It
has long been known that some quantities in certain integrable
boundary field theories (for instance, the current in the boundary
sine-Gordon model \FLSbig, as well as the magnetizations in the Kondo
model \FW) do satisfy a weak to strong coupling duality. Moreover,
inspired by work in supersymmetric gauge theories, it was shown in
\refs{\Paul,\FS,\FSii}\ that these quantities can be expressed, at
$T=0$, in terms of integrals on hyperelliptic curves.  It would be
quite useful to understand the precise mathematical framework
underlying these properties. It seems possible that when combined with
some sort of analyticity and maybe some other type of information,
they can be used to bypass the Bethe ansatz, and lead directly to the
determination of physical quantities in a way analogous to the
supersymmetric gauge theories, as pioneered in \SW.

In this paper, we present some further progress along these lines,
restricting to the simplest case of the boundary sine-Gordon model,
and the underlying $c=1$ conformal field theory. At zero temperature,
we show that it is possible to interpret the fusion relations as
monodromy relations for the current (and magnetizations). This,
combined with duality and an assumption about the number of
singularities in the plane of complex couplings, leads to a direct
determination of physical quantities that indeed bypasses the Bethe
ansatz.

A key ingredient in the following is the current. Although it was
introduced on purely physical grounds in \FLSbig, this quantity turns
out indeed to also be central to our alternative, quite mathematical,
approach: in a nutshell, we will see that the current is the unique
solution of a differential equation that is regular at the origin,
while for instance densities of physical excitations can be expressed
as monodromies around the other singularity (1 or $\infty$).  The
current is so special because it is exactly self-dual. Intuitively,
the property arises because the current is measuring charge transport,
so only operators with charge affect its computation (in contrast with
quantities like the free energy or the magnetizations), and they are
strongly constrained by integrability \LeS.

For convenience, we summarize here the notations to be used in 
the following: 
\eqn\variab{ z=-{(t-1)^{t-1}\over t^t}u^{2(t-1)},
\qquad x\propto z^{-{1\over 2t}}, \qquad u=e^{A+\Delta-\theta}.}
with $\Delta$ defined so that $z(\theta=A)=-1$.

\newsec{Analytic properties of the current}

Since in the case discussed in this paper, we do have the Bethe ansatz
solution available, it is easiest and most concrete to use it to first
illustrate the monodromy properties and the result. In the next
section, we will show how all of these results follow without recourse
to the Bethe ansatz.

\subsec{The model and the current}

The field theory discussed in this paper is usually known as the
boundary sine-Gordon model in its massless limit. It 
consists of a $1+1$-dimensional boson $\phi(x,\tau)$ on the half line
$x\ge 0$: there are no interactions in the bulk and  the action is the
usual $(t/4\pi)\int dx d\tau (\partial_\mu\phi)^2 $, but on the boundary
there is an interaction $\lambda\cos[\phi(0)]$. In addition, there
is a voltage $V$ coupled to the $U(1)$ charge. The dimensionless
couplings in this
problem are $u\propto
V\lambda^{-t/(t-1)}$ (the coefficient of proportionality 
is  non-universal and of little interest here) 
and $t$ (in the notation of \FLSbig\
$u=e^{A+\Delta-\theta_B}$ and $t=1/g$). 
Although physical quantities are continuous in $t$, many of
the formal properties of the system are much easier to understand at
rational or integer values of $t$. This is familiar from the Bethe
ansatz study of the XXZ spin chain, which, with appropriate boundary
conditions, is a lattice discretization of this model.

The interaction violates charge conservation, so one can define the
``current'' describing this charge violation. This situation is
discussed at length in \FLSbig, where integral Bethe ansatz equations
are derived for this current. Here we work with the  normalized quantity 
${\cal I}(u)=tI/V$. At $T=0$, the integral equations become linear, and the
Wiener-Hopf technique allows one to find an integral expression for
the current. This integral expression yields explicit weak-coupling
and strong-coupling series expansions. In terms of the
coupling the expansion of the current for $u$ large, real and positive
(the UV) is
\eqn\uvcurr{{\cal I}(u)=1-\sum_{n=1}^\infty
a_n\left({1\over t}\right)u^{-2n(t-1)/t},}
and for $u$ small (the IR) is
\eqn\ircurr{{\cal I}(u)=\sum_{n=1}^\infty a_n(t) u^{2n(t-1)},}
where
\eqn\coeff{a_n(t)={(-1)^{n+1}\over n!}{\sqrt{\pi}\over 2}{\Gamma(nt+1)\over
\Gamma\left[{3\over 2}+n(t-1)\right]}.}
As noted in \FLSbig, a duality $t\to 1/t$ is already apparent in the
above expressions.
It was shown in \FS\ that these expressions for
the current can be reformulated in terms of  an integral on the hyperelliptic curve
$y^2=x+x^t-u^2$:
\eqn\intrep{{\cal I}(u)=1-{i\over 4u}\int_{{\cal C}} {dx\over y}}
where the contour ${\cal C}$ starts at the origin, loops around the
branch point on the real axis (when $u$ is real), and returns to the
origin. This result is continuous in $t$, but already we see a
difference between $t$ rational or irrational: the curve is of finite
genus when $t$ is rational and infinite genus when it is not.  This
expression means that the current obeys a differential (Picard-Fuchs)
equation, whose order depends on the genus of the curve.

\subsec{Generalized hypergeometric functions}

To discuss the extension of this current to complex couplings, and the
associated analytical properties, it is convenient to consider further
the expansion \ircurr.  For simplicity, we restrict to the simplest
case when $t$ is an integer.
We set 
\eqn\zdef{z=-e^{-2\Delta(t-1)}u^{2(t-1)};}
where
$$\Delta={-1\over 2(t-1)}\left[t\ln t - (t-1)\ln(t-1)\right].$$
The expansion \ircurr\ defines a unique analytic
function for $|z|<1$. Introducing
the generalized hypergeometric function
\eqn\genhyper{{}_pF_q\left[\eqalign{a_1,\ldots,a_p\cr
\rho_1,\ldots,\rho_q\cr};z\right]=\sum_{n=0}^\infty {(a_1)\ldots
(a_p)_n\over (\rho_1)_n\ldots (\rho_q)_n}{z^n\over n!},}
and using the duplication formulas for $\Gamma$ functions, one finds
the simple expression
\eqn\nice{{\cal I}=1-{}_tF_{t-1}\left[\eqalign{
{1\over t}~,~~~~~\ldots~~~~~~~,{t-1\over t},~1~~~\cr
{3\over 2(t-1)},\ldots,{3\over 2(t-1)}+{t-2\over t-1}\cr};
z\right].}
For $t=2$ for instance, we have
\eqn\exampi{\eqalign{
{\cal I}&=1-F\left({1\over 2},1;{3\over 2};z\right)\cr
&=1-{1\over 2z^{1/2}}\ln{1+z^{1/2}\over 1-z^{1/2}}.\cr}}
Like for usual hypergeometric functions, the analytic structure of the
current in terms of the $z$-variable is quite simple. It has two
singular points at $z=1$ and $z=\infty$ , and there is a cut running
from $1$ to $\infty$.  The monodromies around $1$ and infinity are
identical, and the Riemann surface of the function has $t$ sheets
(except for $t=2$ where there are logarithmic terms, as is obvious
from \exampi).

We will write the current for  $|z|>1$ as 
\eqn\dualexp{{\cal I}=1-\sum_{n=1}^\infty  
a_n\left({1\over t}\right)u^{-2n(t-1)/t}=
1-\sum_{n=1}^\infty a_n\left({1\over t}\right)\left(
e^{-i\pi}e^{2(t-1)\Delta} z\right)^{-n/t},}
where $0\leq \hbox{ arg }z<2\pi$.  This expansion, which we obtained
originally from the Bethe ansatz and then from duality considerations,
can also be deduced from standard references on the theory of
generalized hypergeometric functions (Meijer's G functions): see
\ref\Smith{F. Smith, Bull. Am. Math. Soc. (1938) 429;
{\it Higher transcendental functions}, Bateman Manuscript Project
ed.\ by A. Erdelyi, (McGraw-Hill)} for more details. As an example, let
us recall the well-known result for $t=2$:
\eqn\tatai{
\eqalign{1-F\left({1\over 2},1;{3\over 2};z\right)=1- &\Gamma(3/2)\Gamma(1/2)
\left(e^{-i\pi}z\right)^{-1/2} F\left(0,{1\over 2};{1\over 2};z^{-1}\right)\cr
- &{\Gamma(3/2)\Gamma(-1/2)\over \Gamma^2(1/2)}\left(e^{-i\pi}z\right)^{-1}
 F\left({1\over 2},1;{3\over 2};z^{-1}\right),\cr}}
where in fact $F\left(0,{1\over 2};{1\over 2};z^{-1}\right)=1$.
For $t=3$, elementary calculations give:
\eqn\tataii{
\eqalign{1-{}_3F_2&\left[\eqalign{{1\over 3},{2\over 3},1\cr
{3\over 4}~,~{5\over 4}\cr};z\right]=1-c_1\left(e^{-i\pi}z\right)^{1/3}
F\left({7\over 12},{1\over 12}
; {2\over 3};z^{-1}\right)
\cr-
&c_2\left(e^{-i\pi}z\right)^{2/3}F\left({1\over 12},{5\over 12}; {4\over 3}
;z^{-1}\right)
-c_3
\left(e^{-i\pi}z\right)^{-1}{}_3F_2\left[\eqalign{{5\over 4},{3\over 4},1\cr
{5\over 3}~,~{4\over 3}\cr};z^{-1}\right],\cr}}
where the numerical constants are 
\eqn\tataiii{
\eqalign{c_1=&{\Gamma(1/3)\Gamma(3/4)\Gamma(5/4)\over
\Gamma(5/12)\Gamma(11/12)}\cr
c_2=&{\Gamma(-1/3)\Gamma(3/4)\Gamma(5/4)\over \Gamma(1/12)\Gamma(7/12)}\cr
c_3=&{\Gamma(-1/3)\Gamma(-2/3)\Gamma(3/4)\Gamma(5/4)\over
\Gamma(-1/4)\Gamma(1/4)\Gamma(1/3)\Gamma(2/3)}.\cr}}

\subsec{Differential equations}

Using standard results from the theory of generalized hypergeometric
functions, it follows that for $t$ integer, the current solves the
following linear homogeneous differential equation of order $t$
\eqn\geneqdiff{\left[z{d\over dz}\prod_{k=0}^{t-2}\left(z{d\over dz} +{3+2k\over 2(t-1)}
-1\right)-z\prod_{k=1}^{t}\left(z{d\over dz}+{k\over t}\right)\right]
\left(1-{\cal I}\right)=0.}
This can be verified
using either \nice\ or \dualexp. For example, for $t=3$:
\eqn\particuldiffeq{\left[ -{2\over 9}+\left({15\over 16}-{38\over 9}z\right)
{d\over dz}+(3z-5z^2){d^2\over dz^2}+z^2(1-z){d^3\over dz^3}\right]
(1-{\cal I})=0.}

However, the expression \intrep\ of ${\cal I}$ in terms of a hyperelliptic
curve actually requires that ${\cal I}$ obey a differential equation
of order $t-1$ (the Picard-Fuchs equation). This is easy to see:
taking $b$ derivatives with respect to $u$ gives integrals of the form
$$\int_{\cal C} dx\ {x^a\over y^{2b+1}}$$ where $a$ is some integer. If
$a\ge t-1$ or greater, this can be reduced to a sum of integrals with
$a<t-1$ by integrating by parts. 
Explicitly, 
$$\eqalign{\int_{\cal C} {x^a\over y^{2b+1}}&=
{1\over t}\int_{\cal C}dx\ x^{a-t+1}{tx^{t-1}+1\over y^{2b+1}}-
{1\over t}\int_{\cal C}dx\ {x^{a-t+1}\over y^{2b+1}}\cr &=
{2i\delta_{a,t-1}\over t(b-1/2)u^{2b-1}}+ {a-t+1\over
t(b-1/2)}\int_{\cal C}dx\ {x^{a-t}\over y^{2b-1}} - {1\over
t}\int_{\cal C}dx\ {x^{a-t+1}\over y^{2b+1}}\cr}$$ 
The constant term arises because the contour is not closed: the
integration by parts results in a non-vanishing surface term when
$a=t-1$.  By using this relation, one can express the order $t-1$
derivative of ${\cal I}$ as a linear combination (with $u$-dependent
coefficients) of the lower derivatives, so therefore ${\cal I}$
satisfies a differential equation of order $t-1$. Because of the
constant term, this differential equation can have a inhomogeneous piece.

Therefore \geneqdiff\ must be a {\sl total derivative}. This can easily be
checked to be true directly: the reason can be traced back to the fact
that the second product in \geneqdiff\ runs up to $k=t$. Let us
illustrate this for $t=3$: \particuldiffeq\ reduces to
\eqn\particulsimple{\left[{8z\over 9}+{1\over 4}+4z(2z-1){d\over dz}+4z^2(z-1)
{d^2\over dz^2}\right](1-{\cal I})=A_3.}
The value of the constant $A_t$ is determined by the normalization of
${\cal I}$ (here $A_3=1/4$).
The situation is similar for other values of $t$: the current is a
particular solution of an inhomogeneous differential equation of order
$t-1$ which reads
\eqn\simplergeneqdiff{\left[\prod_{k=0}^{t-2}
\left(\delta+{3+2k\over 2(t-1)}-1\right)
-\sum_{k=1}^t b_k {d^{k-1}\over dz^{k-1}} z^k\right](1-{\cal I})=A_t,}
where 
$$b_k={(-1)^k\over k!}\prod_{j=1}^t \left({j\over t}-k-1\right)
-\sum_{j=1}^{k-1} {(-1)^j\over j!}b_{k-j},~~~k=1,\ldots,t-1,~~~b_t=1$$

One obvious question is the significance of the other $t-1$ solutions
of the differential equation \geneqdiff. These solutions are also
necessarily solutions of \simplergeneqdiff\ with $A_t=0$. We will see
in the next section that, at $t$ integer, they coincide with 
the densities of states of the various species of quasiparticles.

Mathematically, it is convenient to express these  other $t-1$ solutions
 in terms of the functions
\eqn\resii{{\cal I}_k(z)= {\cal I}\left[e^{-i\pi(t-k)}z\right]
- {\cal I}\left[e^{i\pi(t-k)}z\right].}
defined for $k=1\dots t-1$ and $|z|>1$.  In other words, we take the
difference of the current on two successive sheets.  The functions
${\cal I}_k$ are defined for $|z|<1$ by first using \resii\ for
$|z|>1$, and then continuing around the singularity. The series
expansions for $z$ large are
\eqn\jbrholden{{\cal I}_k={\sqrt{\pi}
\over it}\sum_{n=1}^\infty {(-1)^{n}\over \Gamma(n)} 
{\Gamma\left({n\over t}\right)\over\Gamma\left[{3\over 2}+n\left({1\over t}-
1\right)\right]}
\sin \left(n\pi{t-k\over t} \right)\left(
e^{-i\pi}e^{2(t-1)\Delta} z\right)^{-n/t}.}
Here it is assumed that $0\leq \hbox{ arg }z<2\pi$.
These coefficients differ from those in the expansion of the current
only by the piece $\sin (n\pi{(t-k)/t})$. Since the solutions of the
differential equation for $|z|$ large are of the form
$$z^{k/t}(a_k(1/t) + a_{k+t}(1/t) z + \dots)$$
it follows immediately that a basis of solutions of the differential
equation consists of the $t$ quantities ${\cal I}(z), {\cal
I}_{t-2p}(z)$ and ${\cal I}_{t-2p-1}\left(e^{i\pi}z\right)\equiv {\cal
I}_{t-2p-1}'(z)$.

For the monodromies to be discussed in the next section, we will also need the
IR expansion valid at small $z$. This does not follow instantly from
\resii, because the relation \resii\ does not hold at small $z$. These
expansions must be determined by continuing the expansions in
\jbrholden\ around the singularity at $z=1$ (this is easiest to do by
writing a contour integral whose residues give the terms in
\jbrholden; this form arises naturally in the Wiener-Hopf
analysis). One finds
\eqn\jbrholdeni{\eqalign{{\cal I}_k=&{i\pi t\over 2(t-1)}m_k \left(
e^{-i\pi}z\right)^{-1/(2t-2)}+{i\sqrt{\pi}\over t-1}
\sum_{n=0}^\infty {(-1)^n\over n+1}\cos\left[\pi\left(n+{1\over 2}\right)
{t-k-1\over t-1}\right]\cr &{\Gamma
\left[-{(2n+1)\over 2(t-1)}\right]
\over \Gamma(n+1)\Gamma \left[-{(2n+1)t\over 2(t-1)}\right] }
\left(e^{-i\pi} e^{2(t-1)\Delta} z\right)^{2n+1\over 2(t-1)},~
k=1,\ldots,t-2\cr}}
where $m_k=2\sin(k\pi/(2t-2))$, $G_+(0)$ and $G_+(i)$ 
are some numerical constants (see \FLSbig\ for their exact values).
The function ${\cal I}_{t-1}$ is given by \jbrholdeni\ for $k=t-1$ with
an additional factor of $1/2$ multiplying the right-hand-side.
 The radius of convergence of
these expansions is the same as for the current: they converge for
$|z|>1$ and $|z|<1$ respectively.

\subsec{$t$ rational}

The situation is rather similar for rational $t$, and depends
mostly upon the value of the numerator $Q$ in $t=Q/P$. Here, we shall
simply give the differential equation
satisfied by the current:
\eqn\newdiffeq{
\left[\prod_{k=0}^{P-1}\left(P\delta-k\right)\prod_{l=0}^{Q-P-1}
\left(\delta+{3+2k\over 2(Q-P)}-1\right)-z\prod_{k=1}^{Q}
\left(\delta+{k\over Q}\right)\right]
(1-{\cal I})=0,}
where now 
\eqn\newz{z=(-1)^P {Q^Q\over (Q-P)^{Q-P}}u^{2(Q-P)}.}
This is a homogeneous differential equation of order $Q$. The fact
that the current can be expressed in terms of the curve
$y^2=x^P+x^Q-u^2$ means that this equation can be reduced to an
inhomogeneous equation of order $Q-1$, just as in the case $t$ integer.

\subsec{Monodromies}

The differential equation \geneqdiff\ is analytic in $z$, so when one
continues a solution around a singularity, one must obtain a solution
upon return to the starting point. However, the continuation of a
given solution is not necessarily the same solution: it only need be
some linear combination of all the solutions.  This linear combination
is called a monodromy. Studying the monodromies is useful for
understanding the solutions; in fact, if one knows all  the
monodromies, then one can reconstruct the differential equation
automatically. In this section, we construct explicitly the monodromies
 of the solutions from the small- and large-$z$
expansions. Later, we will reverse the order of the logic: we will
derive the monodromies from general arguments and therefore infer the
solutions.

The monodromies around infinity give the behavior of the solutions
when we move $z\to e^{2i\pi}z$ for fixed $|z|>1$. Since there are $t$
solutions for the differential equation, the Riemann surface for the
current is $t$ sheeted, with a cut running from $1$ to $\infty$.
The monodromies around infinity for the the whole set of solutions of
the differential equation can be easily written by using the UV
expansion. We have
\eqn\monodromyuv{\eqalign{{\cal I}\left(e^{2i\pi} z\right)=
&{\cal I}-{\cal I}_{t-1}'\cr
{\cal I}_{t-1}'\left(e^{2i\pi}z\right)=&-{\cal I}_{t-2}+{\cal I}'_{t-3}\cr
{\cal I}_{t-2p}\left(e^{2i\pi}z\right)=&{\cal I}'_{t-2p+1}-{\cal I}_{t-2p}+
{\cal I}'_{t-2p+1}\cr
{\cal I}'_{t-2p-1}\left(e^{2i\pi}z\right)=&{\cal I}'_{t-2p+1}-{\cal I}_{t-2p}+
{\cal I}_{t-2p-1}'-{\cal I}_{t-2p-2}
+{\cal I}'_{t-2p-3}\cr
&\ldots\cr}}
The last ``boundary'' terms depend on whether $t$ is odd or even. For
instance, when $t=2$
\eqn\monodrii{\eqalign{{\cal I}\left(e^{2i\pi}z\right)=
&{\cal I}(z)-{\cal I}'_1(z)\cr
{\cal I}'_1\left(e^{2i\pi}z\right)=&-{\cal I}'_1(z)\cr}}
and when $t=3$,
\eqn\monodriii{\eqalign{{\cal I}\left(e^{2i\pi}z\right)=&{\cal I}(z)-{\cal 
I}'_2(z)\cr
{\cal I}_2'\left(e^{2i\pi}z\right)=&-{\cal I}_1(z)\cr
{\cal I}_1\left(e^{2i\pi}z\right)=&{\cal I}_2'(z)-{\cal I}_1(z)\cr}}

The monodromies around the origin are also quickly worked out from the
IR expansions.  The current in particular is regular at $z=0$,
and we  have
\eqn\monodromyir{\eqalign{{\cal I}\left(e^{2i\pi} z\right)=&{\cal I}\cr
{\cal I}_{t-1}'\left(e^{2i\pi}z\right)=&{\cal I}'_{t-1}-{\cal I}_{t-2}+{\cal 
I}'_{t-3}\cr
{\cal I}_{t-2}\left(e^{2i\pi}z\right)=&2{\cal I}'_{t-1}+{\cal I}'_{t-3}-
{\cal I}_{t-2}\cr
{\cal I}_{t-2p}\left(e^{2i\pi}z\right)=&{\cal I}'_{t-2p+1}-{\cal I}_{t-2p}+
{\cal I}'_{t-2p-1}
\cr
{\cal I}_{t-2p-1}\left(e^{2i\pi}z\right)=&{\cal I}'_{t-2p+1}-{\cal I}_{t-2p}+
{\cal I}'_{t-2p-1}-{\cal I}_{t-2p-2}+{\cal I}_{t-2p-3}'\cr
&\ldots\cr}}
The last boundary terms depend on whether $t$ is odd or even. For
instance, when $t=2$
\eqn\monodrii{\eqalign{{\cal I}\left(e^{2i\pi}z\right)=&{\cal I}\cr
{\cal I}'_1\left(e^{2i\pi}z\right)=&-{\cal I}'_1\cr}}
and when $t=3$,
\eqn\monodriii{\eqalign{{\cal I}\left(e^{2i\pi}z\right)=&{\cal I}\cr
{\cal I}_1\left(e^{2i\pi}z\right)=&2{\cal I}_2'-{\cal I}_1\cr
{\cal I}_2'\left(e^{2i\pi}z\right)=&{\cal I}'_2-{\cal I}_1\cr
}}

Finally, the monodromies around $z=1$ follow from the other two,
because going around a circle at large $z$ is equivalent to going
around $z=0$ and $z=1$. Since the monodromies take a solution to a
linear combination of the other solutions, we can represent them
conveniently by matrices $M_\infty$, $M_0$ and $M_1$. The conventions
for orienting the monodromies are usually taken so that one has
$M_\infty=  M_0 M_1$. For instance, one has for $t=3$,
$$
M_1=\left(\eqalign{1~~~ 0 -1\cr
0~~~ 1~~~ 1\cr
0~~~ 0~~~  1\cr}\right)
$$
One checks on this example 
the general property that the monodromies of the current around
$z=1$ and $z=\infty$ are identical, since the current has a trivial
monodromy around the origin. 

%Let us illustrate this explicitly 
%for  $t=2$ by comparing the expression  \exampi\ and the expansion
% \ircurr. Indeed, we see
%that in the region $|z|>1$,
%$$
%{\cal I}(e^{2i\pi}z)-{\cal I}(z)=
%{\sqrt{\pi}\over 2}\sum_{n=1}^\infty (-1)^{n+1}
%{\Gamma(1+n/2)\over \Gamma(n+1)\Gamma(3/2-n/2)}\left({z\over 4}\right)^{-n/2}
%\times -2i\sin{n\pi\over 2}
%$$
%All terms here  vanish except the term $n=1$, giving ${\cal I}\left(
%e^{2i\pi} z\right)-{\cal I}(z)=-{i\pi\over z^{1/2}}$.
%Meanwhile, from \exmapi, one  also finds
%$$
%{\cal I\left(1+e^{2i\pi}(z-1)\right)-{\cal I}(z)=-{i\pi\over z^{1/2}}
%$$

We observe that one has to be careful in performing the continuation
of the ${\cal I}'_{t-2p-1}$ quantities for $|z|<1$, since the
expressions \jbrholden,\jbrholdeni\ hold only for $0\leq \hbox{ arg }
z<2\pi$. But this continuation is easily worked out by using the
previous relations.

A striking feature of the relations \monodromyuv,\monodromyir\ is that
the monodromies for the densities do not involve the current
itself. This is because (or conversely, requires that) they satisfy a
{\it homogeneous} linear differential equation \simplergeneqdiff\ of
order $t-1$, one order less than the hypergeometric equation
\geneqdiff\ we started with. This order $t-1$ differential equation is
the same one that the current satisfies, except that it does not have 
an 
inhomogeneous term.

\newsec{A determination of the current from duality}

In this section we will show that the monodromies derived above from
the explicit Bethe ansatz solution can be obtained directly from
general arguments, at least when $t$ is rational. Therefore, the
explicit solutions can be found without recourse to the Bethe ansatz.

As was discussed at length in \refs{\BLZ,\FLeSii},
the partition functions $Z_j$ of the
spin $j/2$ Kondo model satisfy, as a consequence of the Yang Baxter
equation, some general fusion properties. There are various ways to
write those. A particularly convenient one here is based on the
quantities $Y_k\equiv Z_{k+1}Z_{k-1}$; one has then, in standard notations
\eqn\yfusion{Y_k\left(q^{1/2}x\right)Y_k\left(q^{-1/2} x\right)=
\left[1+Y_{k+1}(x)\right]\left[1+Y_{k-1}(x)\right],}
where $q=e^{i\pi /t}$, and $x\propto \lambda \propto z^{-{1\over 2t}}$
(which is $\propto
e^{(1-g)\theta_B}$ in the notation of \FLSbig). For $t$ an integer, these relations hold only for $k=1,\ldots,t-3$.  Their
closure is a slightly more delicate matter, that was discussed in
detail in section 4 of \FLeSii. It requires the introduction of a pair
of objects, $Y_\pm$, for which one has ${Y_+\over Y_-}= e^{V\over T}$,
and
\eqn\yfusioni{Y_{t-2}\left(q^{1/2}x\right)Y_{t-2}\left(q^{-1/2} x\right)=
\left[1+Y_{t-3}(x)\right]\left[1+Y_+(x)\right]\left[1+Y_-(x)\right],}
together with
\eqn\yfusionii{
Y_+\left(q^{1/2}x\right)Y_+\left(q^{-1/2} x\right)
Y_-\left(q^{1/2}x\right)Y_-\left(q^{-1/2} x\right)=\left[1+Y_{t-2}(x)\right]^2.}
Finally, 
we need  the relation between the current and partition functions
conjectured in \ref\FLSi{P. Fendley, F. Lesage, H. Saleur, J. Stat. Phys. 
79 (1995) 799, hep-th/9409176.}:
\eqn\ifusion{I\left(q^{1/2}x\right)-I\left(q^{-1/2}x\right)=-{i\pi T\over t}
x{\partial\over\partial x}\ln {1+ Y_+\over 1+Y_-}.}
This is true at any temperature $T$.

The claim we are making is that the relations \yfusion,\yfusioni,\yfusionii\
together with \ifusion, and {\sl duality},
are equivalent to the monodromy relations studied previously,
and hence lead to a complete determination of the physical quantities in 
the problem of interest.

General considerations about perturbation near the UV and IR fixed
point show that the quantities $Y$ should have a singularity at $x=0$
and at $x=\infty$. In addition, there should be at least one other
singularity. We have to make the hypothesis there is {\sl only} one
such singularity, which we set at $z=1$ by convention. There are thus
two cuts in the complex $z$ plane for our quantities, one from
$-\infty$ to the origin, and one from $1$ to $\infty$. The fusion
relations are now simply interpreted as monodromy relations. To show
this, we first observe that we can set $Y_k\equiv
\exp{\epsilon_k\over T}$ ($\epsilon_k$ is necessarily positive to guarantee 
that the partition functions are not trivial, even in that limit).
Writing the relations in terms of our usual variable $z$, and continuing
them to the entire complex plane, leads to 
\eqn\betterfusion{\epsilon_k(e^{i\pi}z)+
\epsilon_k(e^{-i\pi}z)=\epsilon_{k-1}(z)+ \epsilon_{k+1}(z)\qquad
k=1,\ldots,t-3}
where we set $\epsilon_0\equiv 0$.
Thus if ${\cal I}_k\propto  z{d\epsilon_k\over dz}$, these
relations are equivalent to all but the first two monodromy relations
in \monodromyuv, \monodromyir.

Next, we set $Y_+=e^{\epsilon_s/T}$ and $Y_-= e^{\epsilon_a/T}$. It is
not clear now what the sign of these $\epsilon$'s is, since the
$Y_\pm$ have no direct physical meaning as partition functions.  To
solve this difficulty, we need {\sl duality}: the latter implies that
the current should expand in powers of $z^n$ in the IR \foot{Recall
that, as discussed in \FS, this follows because the current is
entirely determined by the leading irrelevant operator near the IR
fixed point, unlike the magnetizations or $Y_k$'s.} i.e.\ have no
non-trivial monodromy around the origin. This implies that the right
hand side of \ifusion\ vanishes, and thus, setting
$Y_+=e^{\epsilon_s/T}$ and $Y_-= e^{\epsilon_a/T}$, that both
$\epsilon_s $ and $\epsilon_a$ have a positive real part and are equal
for $|z|<1$. As a result, the fusion relations for $|z|<1$ close with
\eqn\evenbetter{{\cal I}_{t-2}(e^{i\pi}z)+{\cal I}_{t-2}(e^{-i\pi}z)
={\cal I}_{t-3}+2{\cal I}_{t-1},}
where
$${\cal I}_{t-1}\propto {1\over 2}\left(z{d\epsilon_s\over dz} +
z{d\epsilon_a\over dz}\right),~\epsilon_s=\epsilon_a,$$
and the coefficient of proportionality is the same as for 
${\cal I}_k$. No special relation is required for the current this time,
since it has no monodromy around the origin,
\eqn\tutu{{\cal I}(e^{2i\pi z})={\cal I}.}
Finally,  the last relation in \yfusionii\ gives
\eqn\tutui{{\cal I}_{t-1}(e^{i\pi z})+{\cal I}_{t-1}(e^{-i\pi z})
={\cal I}_{t-2}(z).}

In the UV on the other hand, the current does have a non-trivial
monodromy, which implies that the right hand side of \ifusion\ does not
vanish, and $\epsilon_s$ has a negative real part  for  $|z|>1$. Therefore,
for $|z|>1$,  the fusion relations close with
\eqn\evenbetter{{\cal I}_{t-2}(e^{i\pi}z)+{\cal I}_{t-2}(e^{-i\pi}z)=
{\cal I}_{t-3}+ {\cal I}_{t-1},}
where now ${\cal I}_{t-1}\propto z{d\epsilon_s\over dz}$ (still
with the same proportionality coefficient). In addition, \ifusion\ reads
\eqn\tutuii{{\cal I}(e^{i\pi z})-{\cal I}(e^{-i\pi} z)={\cal I}_{t-1}.}

In both cases, it is easy to show that these relations are entirely equivalent
to those of the previous section.

We thus see that all the monodromies are determined by the fusion
relations, together with the relation between the current and the
partition function \ifusion.  Knowing these relations is like knowing
the monodromies, from which the differential equation follows. The
current is then quickly identified as the solution that is regular at
$z=0$, as a result, again, of duality.

\newsec{The other solutions}
\subsec{The other solutions are quasiparticle densities}

The physical significance of the other solutions to the differential
equation is easy to establish when $t$ is integer. The $\epsilon_k$
arise in the Bethe ansatz as the energy it takes to create a single
quasiparticle, and their derivatives correspond to  the densities of
these quasiparticles (or their holes).

These quasiparticles are familiar from the
study of the sine-Gordon model, and consist of a soliton and
antisoliton, along with $t-2$ breathers, which are soliton-antisoliton
bound states. In the presence of the voltage, there are no breathers
in the ground state at zero temperature, i.e. only ``breather
holes''. The energy $\epsilon_k$
required to create a $k^{th}$ breather of mass $m_k=2\sin\left(
{k\pi\over 2( t-1)}\right)$ can be computed by the same technique as
the one used to compute the current. The density of holes is simply
related by $2\pi
\rho_k^h(\theta)={d\epsilon_k
\over d \theta}\propto z{\partial \epsilon_k\over\partial z}$, where,
in the notations of \FLSbig, the rapidity $\theta$ of the particles is
defined by the requirement that $\epsilon_k\propto m_k e^\theta$
at large $\theta$. It is related to $u$ via $u=e^{A+\Delta-\theta}$, 
$A$ is the Fermi
rapidity; $z$ is still related with $u$
by \zdef, the UV region
corresponds to $Re(\theta)>A$, the IR to $Re(\theta)<A$. 
The densities are precisely related to the ${\cal I}_k$ we introduced 
earlier through  the
normalization
\eqn\dimless{2\pi\rho_k^h = {(t-1) V \over i\pi t}{\cal I}_k 
\qquad k=1\dots t-1.}
Here $V$, the physical voltage, is proportional to $e^A$
(see \FLSbig\ for the exact value of this coefficient).

For $k=t-1$, the meaning of formula \dimless\ is that
$2\pi \rho_{t-1}^h=2\pi\rho_a^h$. In fact, for $\theta>A$, $\rho_s=0$,
but for $\theta<A$, $\rho_s=\rho_a^h$, so formula \dimless\ gives
this density too.

For $t=2$, the particles do not interact, so there are no
breathers. The density of states for free particles means that simply
$2\pi\rho^h_a=e^\theta$, as is easy to establish for a free
theory. In terms of the variable $z$, this goes as ${1\over
\sqrt{z}}$, which is precisely the second solution
$z^{-1/2}F\left(0,{1\over 2};{1\over 2};z\right)$ of the
hypergeometric equation corresponding to \exampi, as illustrated in
\tatai.
This agrees with the series expansions: from \exampi, we see
that in the region $|z|>1$,
$$
{\cal I}(e^{2i\pi}z)-{\cal I}(z)=
{\sqrt{\pi}}
\sum_{n=1}^\infty (-1)^{n}i\sin\left({n\pi\over 2}\right)
{\Gamma(1+n/2)\over \Gamma(n+1)\Gamma(3/2-n/2)}\left({z\over 4}\right)^{-n/2}
$$
All terms here vanish except the term $n=1$, giving ${\cal
I}\left(e^{2i\pi} z\right)-{\cal I}(z)=-{i\pi\over z^{1/2}}$, and thus
${\cal I}_1(z)=-{\pi\over z^{1/2}}$. Using that $e^A={V\over2},
e^\Delta={1\over 2}$ for $t=2$, this gives rise to
$2\pi\rho_1^h=e^\theta$, as expected.

The fact that the ${\cal I}_k$ are related to quasiparticle densities
gives a physical significance to the radius of convergence for the
expansions. In terms of $\theta$, the circle of convergence $|z|=1$
intersects the real axis for $\theta=A$. This follows from the fact
that $A$ is the Fermi rapidity, where some quantities are expected to
be singular; for instance, the density for solitons vanishes for
$\theta>A$ but is non zero for $\theta<A$.

Notice that, unlike the current, the
other solutions have a singularity at the origin as well as at
$z=1$ and $z=\infty$. This has a simple physical
interpretation in terms of the operators controlling the behavior of
these quantities near the IR fixed point. As mentioned in the
introduction, the current is
controlled only by the tunneling operator $\cos t
\tilde{\Phi}$ ($\tilde{\Phi}$ the dual boson); the densities, like thermal properties, depend also on
the local conserved quantities which are polynomials in the
derivatives of $\tilde{\Phi}$ \LeS.

For completeness, we finally would like to give the expansions for the
quantities $Z_{BSG}$ of \FLeSii; one has
\eqn\details{\eqalign{T {\partial\over\partial \theta}
&\ln Z_{BSG}(z,\pm iV/2T)=\cr
&\pm V{t-1
\over 2t^2\sqrt{\pi}}\sum_{n=1}^\infty {(-1)^{n+1}\over \Gamma(n)} 
{\Gamma\left({n\over t}\right)\over\Gamma\left[{3\over 2}+n\left({1\over t}-
1\right)\right]}
{e^{\pm ni\pi/t}\over 2i\cos {n\pi\over t}}\left(
e^{-i\pi}e^{2(t-1)\Delta} z\right)^{-n/t}.\cr}}

\subsec{Representing the densities with the curve}

The ${\cal I}_k$ and hence the densities can also be represented by
integrals on the same hyperelliptic curve as the current in \intrep.
This is very similar (and sometimes identical) to what happens for the
higher-spin magnetizations in the Kondo problem, as discussed in
\refs{\Paul,\FSii}. For simplicity, we will discuss the case $t$ odd
only, for which it is convenient to make a change of variable $x\to
{1\over x}$, so the current can be written
\eqn\intrepi{1-{\cal I}={1\over 4iu}\int_{{\cal C}_0} { x^{(t-3)/2} dx\over 
\sqrt{x^{t-1}+1-u^2 x^t}}\equiv \int_{{\cal C}_0}{dx\over \widetilde y}.}
Here the contour starts at infinity, loops around the branch point on
the positive real axis, and goes back to infinity.

In general, the integrand has $t$ branch points, which, for large $u$
are approximately located at $x_k=u^{-2/t} e^{2i\pi k/t}, k=0, \ldots,
t-1$.  One branch point is on the positive real axis, and at large $u$
the others are symmetrically distributed in the complex plane. As a
function of $u$, the current has singularities when $z=1$, an equation
with $2(t-1)$ solutions.  As $u$ loops around one of the
singularities, $z$ loops around $1$, and monodromies are picked up.
Of course, as $u$ goes around one of the singularities, the different
branch points of the integrand are exchanged. As a result, or, more
simply, by using formula \resii, one can show the following. The
dimensionless reduced densities for breathers with $t-k$ even can be
written simply as
\eqn\intdens{{\cal I}_{t-2p}(z)=\int_{{\cal C}_{p,t-p}} {dx\over 
\widetilde y},}
where ${\cal C}_{p,t-p}$ loops around the singularities $x_{p}$ and
$x_{t-p}$, i.e.
 the $(p+1)^{th}$ cycle of the Riemann surface of the
integrand, as shown on the figure ( ${\cal C}_0\equiv {\cal C}_t$ can
be considered the first cycle). For $g={1\over 3}$ for instance, we
thus see that the integral around one cycle is ${\cal I}_1$, the
density of the first breather, while the integral around the second
cycle is the current.

\fig{A choice of cuts and contours in the complex $x$ plane for $t=5$. The integral
around ${\cal C}_0$ gives the current, while integrals around ${\cal C}_{23}$ (resp. ${\cal C}_{14}$) give ${\cal I}_1$ (resp. ${\cal I}_3$).}{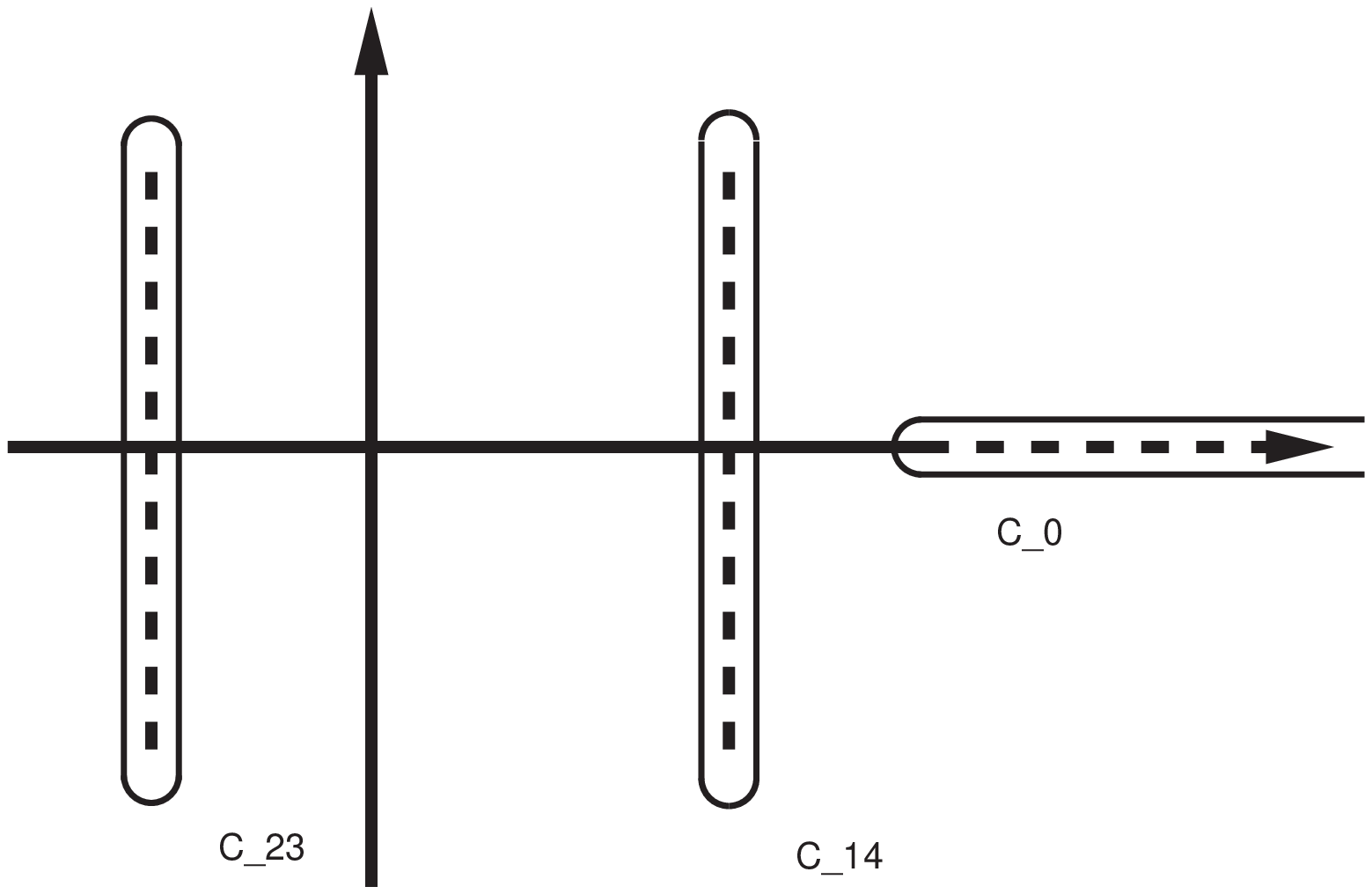}{8cm}
\figlabel\tabb

It is also possible to express the dimensionless densities for
breathers with $t-k$ odd by choosing different cycles. For instance, by
writing
$$
2i\sin {n\pi\over t}= e^{-ni\pi/t}\left(e^{2ni\pi/t}-1\right)
$$
one can show that ${\cal I}_{t-1}\left(e^{i\pi} z\right)$ reads as
\intdens\ but with the integral taken along a contour ${\cal C}_{10}$
that loops around $x_1$ and $x_0$. More generally, one has
\eqn\intdensi{{\cal I}'_{t-2p-1}(z)={\cal I}_{t-2p-1}\left(e^{i\pi}z\right)=
\int_{{\cal C}_{p,p+1}} {dx\over \widetilde y},}
as illustrated in 
figure 2. 

\fig{An other  choice of cuts and contours in the complex $x$ plane for $t=5$. The integral
around ${\cal C}_0$ gives the current, while integrals around ${\cal C}_{12}$ (resp. ${\cal C}_{34}$) give ${\cal I}_2'$ (resp. $\bar{{\cal I}}_2'$).}{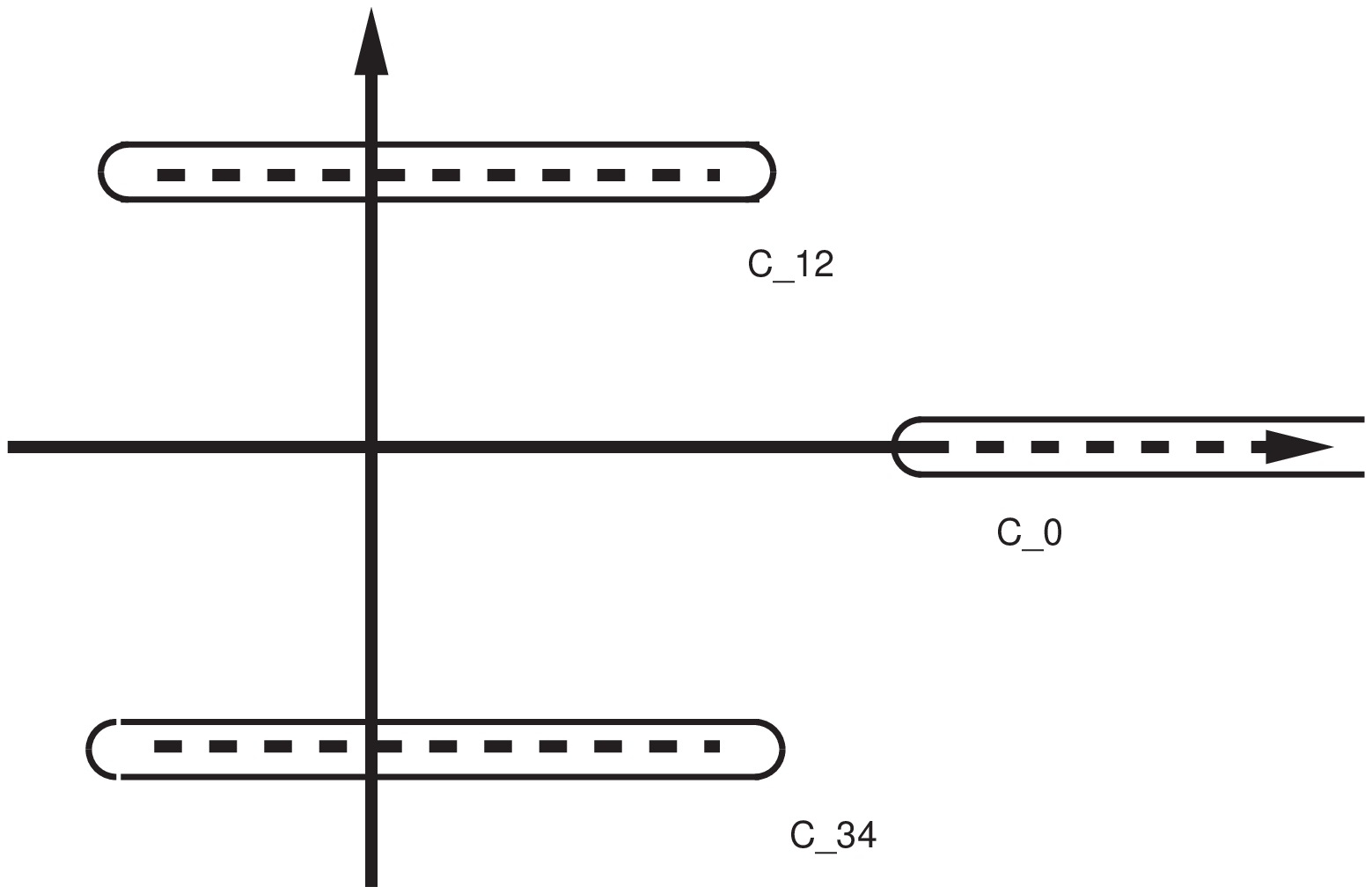}{8cm}
\figlabel\tabb

Similarly, ${\cal
I}_{t-2p-1}\left(e^{-i\pi} z\right)$ reads as \intdensi, but with a
complex conjugate contour. As a result, fusion relations take a very
simple form.  For instance, one has ${\cal I}_{t-3}\left(e^{i\pi}
z\right)+{\cal I}_{t-3}\left(e^{-i\pi}z\right)= {\cal
I}_{t-2}(z)+{\cal I}_{t-4}(z)$, is illustrated on figure 3. 

\fig{Graphical representation of the fusion relation
${\cal I}_{2}\left(e^{i\pi}
z\right)+{\cal I}_{2}\left(e^{-i\pi}z\right)= {\cal
I}_{3}(z)+{\cal I}_{1}(z)$ for $t=5$.}{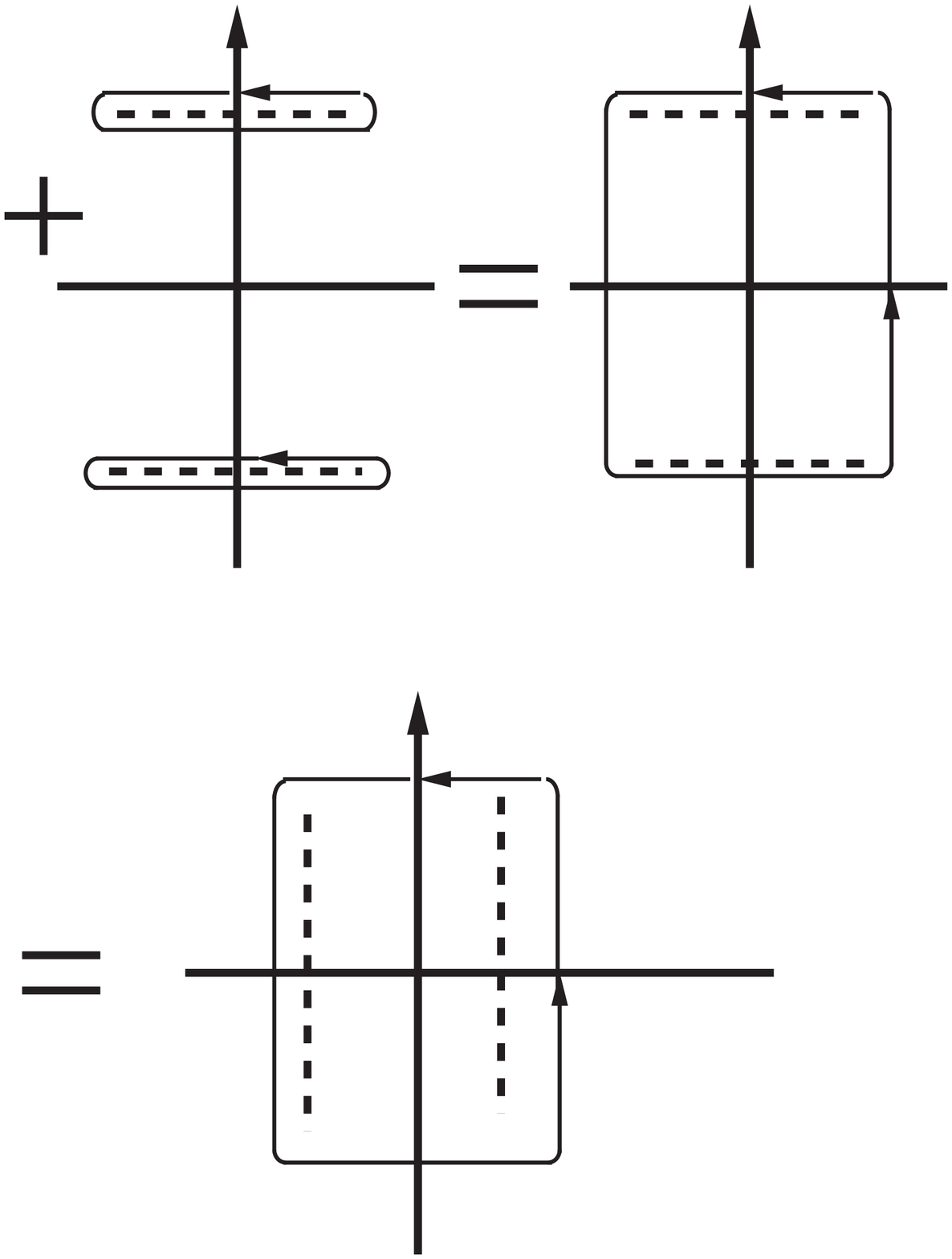}{8cm}
\figlabel\tabb

Also,
the truncation of the fusion relations ${\cal I}_{t-1}\left(e^{i\pi}
z\right)+ {\cal I}_{t-1}\left(e^{-i\pi}z\right)= {\cal I}_{t-2}(z)$
can be illustrated graphically in a similar fashion.

To conclude, we illustrate these considerations by the case $t=2$. We
now have only two branch points, and the integral of the differential
form along the unique cycle is (going back to the original $x$
variable)
$$
2{i\over 4}\int_{x_0}^{x_1}{dx\over u\sqrt{x+x^2-u^2}}=-{\pi\over 2u},
$$
where $x_0$ and $x_1$ are the two branch points. On the other hand, 
$${\cal I}_1(z)={2i\pi\over V}{2\pi\rho_a^h}= {2i\pi\over
V}e^{\theta}={i\pi\over 2u},$$
so ${\cal I}_1\left(e^{i\pi} z\right)={-\pi/(2u)}$ as required.

\newsec{Conclusion and Questions}

This paper we hope is a first step in finding new ways to solving
integrable models.  The case we have discussed can be thought of as
the deformed $SU(2)_1$ case (when $t=1$ the bulk conformal field
theory is the $SU(2)_1$ WZW model). This has been well understood by
many approaches, and so our results do shed light on the picture but
don't result in the computation of any new quantities. In the
$SU(2)_k$ case (the multi-channel quantum wire and multi-channel Kondo
problem), the zero-temperature curves \FSii\ are known and the
multi-channel Kondo problem fairly well understood (see \ref\Cox{
D. Cox and A. Zawadowski, Adv. Phys. 47 (1998) 599,
cond-mat/9704103} and references within). The functional relations can
easily be derived for most of the quantities, but the crucial one
involving the current (the analog of $\ifusion$) is not yet known. In
the $SU(N)$ case,, the problem is even murkier. The
$SU(N)$ Kondo problem is integrable \Kondorev, but only in a special
case has the zero-temperature magnetization been calculated. Moreover,
it is not even known which if any $SU(N)$ generalizations of the boundary
sine-Gordon problem are integrable. We hope that  our approach will be 
useful in understanding these problems.

More formally, observe that, although we were motivated by the study of the 
boundary sine-Gordon model, all the results obtained 
do have a well defined meaning within the $c=1$ conformal field theory itself,
and its integrable structure. It is tempting to speculate 
that monodromy and duality should play a crucial role in the study of this
integrable structure, in particular that there should be hyperelliptic curves
``naturally'' associated with rational conformal field theories. More progress
in understanding the general picture is clearly needed here.

Finally, we find it intriguing that  the curves are so simple and
continuous in $t$, given that the differential equations change
dramatically as $t$ is varied. It is hard not to believe that there is
a simple and deep reason underlying this, but we do not know what this
is.

\bigskip\bigskip

This research was supported by the NSF (DMR-9802813), the DOE
(Outstanding Junior Investigator award) and by the Sloan Foundation (P.F.);
and by the NYI program 
(NSF-PHY-9357207)
and DOE grant DE-FG03-84ER40168 (H.S.).

%\ifx\answ\bigans\vfill\eject\fi

\listrefs

\bye